\documentclass[12pt]{article}
\usepackage{fancyhdr}

\usepackage{mathrsfs}
\usepackage[T1]{fontenc}
\usepackage{mathpazo}
\usepackage{setspace}
\usepackage{amsfonts}
\usepackage{amssymb}
\usepackage{amsmath}
\usepackage{epsfig}
\usepackage{latexsym}
\usepackage{color}
\usepackage{graphicx}
\usepackage{nicefrac}
\usepackage[latin1]{inputenc}
\usepackage{slashed}
\usepackage{multirow}
\usepackage{comment}
\usepackage{soul}
\usepackage{hyperref}
\usepackage[nosort]{cite}
\usepackage{datetime}
\usepackage{braket}

\usepackage[titletoc]{appendix}

\def\hybrid{\topmargin -20pt    \oddsidemargin 0pt
        \headheight 0pt \headsep 0pt
        \textwidth 6.25in       
        \textheight 9.25in       
        \marginparwidth .875in
        \parskip 5pt plus 1pt   \jot = 1.5ex}

\hybrid

\def\baselinestretch{1.2}

\catcode`\@=11

\def\marginnote#1{}
\newcount\hour
\newcount\minute
\newtoks\amorpm
\hour=\time\divide\hour by60
\minute=\time{\multiply\hour by60 \global\advance\minute by-\hour}
\edef\standardtime{{\ifnum\hour<12 \global\amorpm={am}%
        \else\global\amorpm={pm}\advance\hour by-12 \fi
        \ifnum\hour=0 \hour=12 \fi
        \number\hour:\ifnum\minute<10 0\fi\number\minute\the\amorpm}}
\edef\militarytime{\number\hour:\ifnum\minute<10 0\fi\number\minute}

\def\draftlabel#1{{\@bsphack\if@filesw {\let\thepage\relax
   \xdef\@gtempa{\write\@auxout{\string
      \newlabel{#1}{{\@currentlabel}{\thepage}}}}}\@gtempa
   \if@nobreak \ifvmode\nobreak\fi\fi\fi\@esphack}
        \gdef\@eqnlabel{#1}}
\def\@eqnlabel{}
\def\@vacuum{}
\def\draftmarginnote#1{\marginpar{\raggedright\scriptsize\tt#1}}

\def\draft{\oddsidemargin -.5truein
        \def\@oddfoot{\sl preliminary draft \hfil
        \rm\thepage\hfil\sl\today\quad\militarytime}
        \let\@evenfoot\@oddfoot \overfullrule 3pt
        \let\label=\draftlabel
        \let\marginnote=\draftmarginnote
   \def\@eqnnum{(\theequation)\rlap{\kern\marginparsep\tt\@eqnlabel}%
\global\let\@eqnlabel\@vacuum}  }

\def\preprint{\twocolumn\sloppy\flushbottom\parindent 2em
        \leftmargini 2em\leftmarginv .5em\leftmarginvi .5em
        \oddsidemargin -.5in    \evensidemargin -.5in
        \columnsep .4in \footheight 0pt
        \textwidth 10.in        \topmargin  -.4in
        \headheight 12pt \topskip .4in
        \textheight 6.9in \footskip 0pt
        \def\@oddhead{\thepage\hfil\addtocounter{page}{1}\thepage}
        \let\@evenhead\@oddhead \def\@oddfoot{} \def\@evenfoot{} }

\def\numberbysection{\@addtoreset{equation}{section}
        \def\theequation{\thesection.\arabic{equation}}}

\def\underline#1{\relax\ifmmode\@@underline#1\else
        $\@@underline{\hbox{#1}}$\relax\fi}

\def\titlepage{\@restonecolfalse\if@twocolumn\@restonecoltrue\onecolumn
     \else \newpage \fi \thispagestyle{empty}\c@page\z@
        \def\thefootnote{\fnsymbol{footnote}} }

\def\endtitlepage{\if@restonecol\twocolumn \else \newpage \fi
        \def\thefootnote{\arabic{footnote}}
        \setcounter{footnote}{0}}  

\catcode`@=12
\relax

\def\figcap{\section*{Figure Captions\markboth
        {FIGURECAPTIONS}{FIGURECAPTIONS}}\list
        {Figure \arabic{enumi}:\hfill}{\settowidth\labelwidth{Figure
999:}
        \leftmargin\labelwidth
        \advance\leftmargin\labelsep\usecounter{enumi}}}
 \relax
\def\tablecap{\section*{Table Captions\markboth
        {TABLECAPTIONS}{TABLECAPTIONS}}\list
        {Table \arabic{enumi}:\hfill}{\settowidth\labelwidth{Table
999:}
        \leftmargin\labelwidth
        \advance\leftmargin\labelsep\usecounter{enumi}}}
 \relax
\def\reflist{\section*{References\markboth
        {REFLIST}{REFLIST}}\list
        {[\arabic{enumi}]\hfill}{\settowidth\labelwidth{[999]}
        \leftmargin\labelwidth
        \advance\leftmargin\labelsep\usecounter{enumi}}}
 \relax
\makeatletter
\newcounter{pubctr}
\def\publist{\@ifnextchar[{\@publist}{\@@publist}}
\def\@publist[#1]{\list
        {[\arabic{pubctr}]\hfill}{\settowidth\labelwidth{[999]}
        \leftmargin\labelwidth
        \advance\leftmargin\labelsep
        \@nmbrlisttrue\def\@listctr{pubctr}
        \setcounter{pubctr}{#1}\addtocounter{pubctr}{-1}}}
\def\@@publist{\list
        {[\arabic{pubctr}]\hfill}{\settowidth\labelwidth{[999]}
        \leftmargin\labelwidth
        \advance\leftmargin\labelsep
        \@nmbrlisttrue\def\@listctr{pubctr}}}
 \relax
\makeatother
\newskip\humongous \humongous=0pt plus 1000pt minus 1000pt

\newif\ifdtup

\relax

\def\be{\begin{equation}}
\def\ee{\end{equation}}
\def\ba{\begin{eqnarray}}
\def\ea{\end{eqnarray}}

\def\del{\partial}

\def\k{\kappa}

\def\a{\alpha}

\def\b{\beta}

\def\g{\gamma}

\def\d{\delta}

\def\l{\lambda}
\def\L{\Lambda}
\def\s{\sigma}

\def\no{\noindent}

\def\qq{\qquad}

\def\IR{\relax{\rm I\kern-.18em R}}

\def \ov {\over}

\def\IR{\relax{\rm I\kern-.18em R}}
\def\IL{\relax{\rm I\kern-.18em L}}

\def\inv{^{\raise.15ex\hbox{${\scriptscriptstyle -}$}\kern-.05em 1}}

\def\Tr{{\rm Tr}}

\begin{document}

\renewcommand{\theequation}{\thesection.\arabic{equation}}
\csname @addtoreset\endcsname{equation}{section}

\newcommand{\beq}{\begin{equation}}
\newcommand{\eeq}[1]{\label{#1}\end{equation}}
\newcommand{\ber}{\begin{equation}}
\newcommand{\eer}[1]{\label{#1}\end{equation}}
\newcommand{\eqn}[1]{(\ref{#1})}
\begin{titlepage}
\begin{center}

\hfill  CERN-TH-2019-186
\vskip .4 cm

\vskip .3 in

{\large\bf Strong integrability of $\lambda$-deformed models}

\vskip 0.4in
{\bf George Georgiou},$^a$\ \!
{\bf Konstantinos Sfetsos}$^a$\ \! and {\bf Konstantinos Siampos}$^{a,b}$
\vskip 0.15in

{\em${}^a$Department of Nuclear and Particle Physics,\\ Faculty of Physics, 
National and Kapodistrian University of Athens,\\15784 Athens, Greece
}

\vskip 0.1in

{\em${}^b$Theoretical Physics Department, CERN, 1211 Geneva 23, Switzerland}

\vskip 0.1in

{\footnotesize \texttt george.georgiou, ksfetsos, konstantinos.siampos@phys.uoa.gr}


\vskip .5in
\end{center}

\centerline{\bf Abstract}

\no
We study the notion of strong integrability for classically integrable $\lambda$-deformed CFTs and coset CFTs.
To achieve this goal we employ the Poisson brackets of the spatial Lax matrix which we 
prove that it assumes the Maillet $r/s$-matrix algebra. As a consequence the 
system in question are  integrable in the strong sense. Furthermore, we show that the derived Maillet $r/s$-matrix 
algebras can be realized in terms of twist functions, at the poles of which we recover the underlying symmetry algebras.

\noindent

\no

\vskip .4in
\noindent
\end{titlepage}
\vfill
\eject

\newpage

\tableofcontents

\noindent

\def\baselinestretch{1.2}
\baselineskip 20 pt
\noindent


\setcounter{equation}{0}
\renewcommand{\theequation}{\thesection.\arabic{equation}}

\section{Introduction and conclusions}

A two-dimensional $\s$-model is integrable provided that its equations of motion can be packed into a flat Lax connection\footnote{
The worldsheet coordinates $\s^\pm$ and $(\tau,\sigma)$ are related by
\begin{equation*}
\s^\pm=\tau\pm\s\,,\quad\del_0=\del_++\del_-\,,\quad\del_1=\del_+-\del_-\,
\end{equation*}
and $\ast\text{d}\s^\pm=\pm\text{d}\s^\pm$ or $\ast\text{d}\s=\text{d}\tau, \ast\text{d}\tau=\text{d}\s$ in Lorentzian signature.
}
\be
\label{Laxeom}
\partial_+ L_--\partial_- L_+=[L_+,L_-]\,,\quad L_\pm=L_\pm(\tau,\sigma;z)\,,\quad z\in\mathbb{C}\,,
\ee
where $z$ is a spectral parameter and $L_\pm$ are the Lie algebra valued $L_\pm=L_\pm^At_A$ Lax matrices. The 
$t_A$'s are the generators of a semi-simple Lie algebra 
satisfying the following normalization and commutation relations 
\begin{equation*}
\text{Tr}\left(t_At_B\right)=\d_{AB}\,,\quad [t_A,t_B]=f_{ABC}\,t_C\, ,
\end{equation*}
where the structure constants $f_{ABC}$ are purely imaginary.  
Using \eqref{Laxeom}, we can design the monodromy matrix 
\begin{equation*}
M(z)=\text{Pexp}\int_{-\infty}^\infty\text{d}\sigma\, L_1\,,\quad L_1=L_+-L_-\,,
\end{equation*}
which is conserved for all values of $z$, i.e. $\partial_0 M(z)=0$. Expanding the monodromy matrix introduces infinite conserved charges 
which a priori are not in involution.

Following \cite{Sklyanin:1980ij}, we compute the equal time Poisson brackets of $L_1$
\be
\label{Sklyanin}
\{L_1^{(1)}(\sigma_1;z),L_1^{(2)}(\sigma_2;w)\}_\text{P.B.}=\{L_1^A(\sigma_1;z),L_1^B(\sigma_2;w)\}_\text{P.B.}\,t_A\otimes t_B\,,
\ee
where the superscript in parenthesis specifies the vector spaces on which the matrices act.\footnote{
In particular
\begin{equation*}
\begin{split}
&N^{(1)}=N_A t_A\otimes\mathbb{1}\,,\quad N^{(2)}=N_A\mathbb{1}\otimes t_A\\
&n^{(12)}=n_{AB} t_A\otimes t_B\otimes\mathbb{1}\,,\quad 
n^{(23)}=n_{AB} \mathbb{1}\otimes t_A\otimes t_B\,,\quad 
n^{(13)}=n_{AB} t_A\otimes\mathbb{1}\otimes t_B\,,
\end{split}
\end{equation*}
for arbitrary vector $N=N_A t_A$ and tensor $n=n_{AB} t_A\otimes t_B$. 
}
It was proven in \cite{Maillet:1985ek,Maillet:1985ec}, that the conserved charges are in involution provided that the brackets assume the $r/s$-Maillet form 
\be
\label{rsMaillet}
\{L_1^{(1)}(\sigma_1;z),L_1^{(2)}(\sigma_2;w)\}_\text{P.B.}=\left([r_{-zw},L_1^{(1)}(\s_1;z)]+[r_{+zw},L_1^{(2)}(\s_2;w)]\right)\d_{12}-2s_{zw}\d'_{12}\,,
\ee
where $\d_{12}=\d(\s_1-\s_2), \d'_{12}=\del_{\s_1}\d(\s_1-\s_2)$ correspond to ultralocal and non-ultralocal terms, respectively. 
The $r_{\pm zw}=r_{zw}\pm s_{zw}$ are matrices on the basis $t_A\otimes t_B$ which depend on $(z,w)$ and as a consequence of the anti-symmetry of the Poisson brackets, it follows that $r_{+zw}+r_{-wz}=0$. A sufficient condition for the Poisson structure to satisfy the Jacobi identity is the (non-dynamical) modified Yang--Baxter (mYB) relation
\be
\label{mYB}
[r_{+z_1z_3}^{(13)},r_{-z_1z_2}^{(12)}]+[r_{+z_2z_3}^{(23)},r_{+z_1z_2}^{(12)}]+[r_{+z_2z_3}^{(23)},r_{+z_1z_3}^{(13)}]=0\,.
\ee
A consequence of the non-ultralocal terms is that the Poisson brackets 
for the monodromy matrix are not well defined. Nevertheless, it is possible 
to regularize the Poiss\-on brackets by using a generalized symmetric limit procedure \cite{Maillet:1985ek,Maillet:1985ec}
arising at
\begin{equation*}
\{M^{(1)}(z), M^{(2)}(w)\}_\text{P.B.}\, = \, [r_{zw}, M(z) \otimes M(w)]
+ M^{(1)}(z)\, s_{zw} M^{(2)}(w)- M^{(2)}(w) \, s_{zw} \, M^{(1)}(z) \,.
\end{equation*}
Let us also note that the vanishing of the Poisson brackets of $\text{Tr}\left(M^m(z_1)\right)$ with 
$\text{Tr}\left(M^n(z_2)\right)$ ($m,n$ are positive integers) is independent of the regularization scheme  \cite{Maillet:1985ek,Maillet:1985ec}.

In this work we focus on the strong integrability of a class of two-dimensional inte\-grable $\sigma$-models, known as $\l$-deformations.  The prototype class of integrable deformations was constructed in \cite{Sfetsos:2013wia}
for  semi-simple group and coset spaces;  the $SU(2)$ example was first found in \cite{Balog:1993es}.
This work was further developed for symmetric coset spaces in \cite{Hollowood:2014rla} and generalized to semi-symmetric coset spaces and applied to integrable deformations of the $\text{AdS}_5\times S^5$ superstring  in \cite{Hollowood:2014qma}. Broadly speaking, this class of integrable deformations 
describes perturbations of WZW or gauged WZW models with current or parafermion bilinears, on group or symmetric cosets, respectively. 
Our present focus is to study their Maillet algebra and to prove that the conserved charges are in involution. 
The assumed $r/s$-Maillet form \eqref{rsMaillet}, can be expressed in terms of a meromorphic function, known as a twist function, at the poles  of which one unveils the underlying symmetry algebra. Finally, the aforeme\-ntioned $r/s$-Maillet form is the first step towards the quantization of this class of integrable deformations. These serve as the main motivations for the present work.

The plan of the paper is as follows:
In Sec. \ref{Maillet.group}, we work out the strong integrability for group spaces. We initiate our study by revisiting a known example, namely the 
isotropic $\l$-deformed model (\S~\ref{isotropic.group}). Next we focus on $\l$-deformations of two current algebras at unequal levels (\S~\ref{kdhksjsjjksn}).
In Sec. \ref{coset.strong.section}, we shift gears and we consider the strong integrability for coset spaces. Contrary to the group spaces the $r/s$-Maillet form is 
sa\-tisfied in the weak sense as we need to include first class constraints which are related to the residual gauge invariance of the subgroup.
At first we revisit the symmetric coset case in (\S~\ref{iso.symmetric.coset}) and then we move on to the $\l$-deformed $G_{k_1}\times G_{k_2}/G_{k_1+k_2}$ 
(\S~\ref{diagonal.symmetric.coset}). Finally in Sec. \ref{more.group}, we provide the 
 $r/s$-Maillet form of the most general $\l$-deformed models consisting of $n$ commuting copies current algebras, at different levels. 
 
The goal of this paper is to study the notion strong integrability for classically integrable $\l$-deformed CFTs and coset CFTs. To achieve this goal
we prove that the spatial Lax realizes the $r/s$-Maillet form \cite{Maillet:1985ek,Maillet:1985ec}. In the coset cases,
we have  two choices to attack the problem. One may either employ the first class constraints, related to the residual gauge symmetry in the subgroup,
or use second class constraints by strongly imposing the constraints and projecting to the coset. In this work we employed the first class constraints. It would be very interesting to study the coset cases by imposing the second class constraints. This computation would require an extension of the $r/s$-Maillet form \eqref{rsMaillet}, in order to also include the antisymmetric step function -- resulting from a Dirac analysis of the second class constraints.

\section{Group spaces}
\label{Maillet.group}

In this section we work out the $r/s$-Maillet form for the isotropic group case at equal and unequal levels. 
In the equal level case, we review the construction in \cite{Itsios:2014vfa} as a warm up for the unequal level case.

\subsection{The isotropic group space}
\label{isotropic.group}

We first revisit a known  example where the above formalism has been already applied in detail. 
We consider the isotropic $\l$-deformed model which is known to be integrable, with action given by \cite{Sfetsos:2013wia}
\be
\label{klfjslkskdjdw}
S_{k,\l}=S_{\text{WZW},k}(\frak{g})+\frac{k}{\pi}\int_{\del B}\text{d}^2\s\,\text{Tr}\left[J_+\left(\lambda^{-1}-D^T\right)^{-1}J_-\right]\,,
\ee
where $S_{\text{WZW},k}(\frak{g})$ is the WZW model at level $k$ of a group element $\frak{g}\in G$
\be
S_{\text{WZW},k}(\frak{g}) = {k\ov 2\pi} \int_{\del B} \text{d}^2\s\, \text{Tr}[\del_+ \frak{g}^{-1} \del_- \frak{g}]+
{k\ov 12\pi} \int_\text{B}  \text{Tr}[\frak{g}^{-1} \text{d}\frak{g}]^3\,,
\ee
and
\be
\label{skfkduedue}
J_{+}^A=-i\,\text{Tr}\left[t_A\del_+\frak{g}\frak{g}^{-1}\right]\,,\quad J_{-}^A=-i\,\text{Tr}\left[t_A\frak{g}^{-1}\del_-\frak{g}\right]\,,\quad 
D_{AB}=\text{Tr}\left[t_A\frak{g}t_B\frak{g}^{-1}\right]\,.
\ee
For an isotropic coupling, namely $\lambda_{AB}=\l\,\d_{AB}$, the equations of motion of \eqref{klfjslkskdjdw} 
take the form \cite{Sfetsos:2013wia,Hollowood:2014rla}
\be
\del_\pm A_\mp=\pm\frac{1}{1+\l}\,[A_+,A_-]\,,
\ee
where 
\be
A_+=i\left(\lambda^{-1}\mathbb{1}-D\right)^{-1}J_+\,,\quad A_-=-i\left(\lambda^{-1}\mathbb{1}-D^T\right)^{-1}J_-\,.
\ee
The above system of equations admits a flat Lax connection \eqref{Laxeom} with
\begin{equation}
L_\pm=\frac{2}{1+\lambda}\frac{z}{z\mp1}A_\pm\,,\quad z\in\mathbb{C}\,.
\end{equation}
Using the above we can build the spatial Lax matrix 
\be
\label{jrkdifyj}
L_1=-\frac{2z}{(1+\l)(1-z^2)}\big((1-z)A_-+(1+z)A_+\big)\,.
\ee
To evaluate \eqref{Sklyanin}, we need the Poisson brackets of $A_\pm$. These can be evaluated by expressing the gauge fields in terms the currents $J_\pm$ and by using  two commuting copies of the current algebra \cite{Rajeev:1988hq,Bowcock:1988xr}
\be
\begin{split}
\label{KM}
& \{J_\pm^A(\s_1),J_\pm^B(\s_2)\}_\text{P.B.}=\frac{2}{k}\left(f_{ABC}J_\pm^C(\s_2)\d_{12}\pm\d_{AB}\d'_{12}\right)\,,
\\
& 
\{J_\pm^A(\s_1),J_\mp^B(\s_2)\}_\text{P.B.}=0\,,
\end{split}
\ee
where  the currents $J_\pm$ are given in terms of $A_\pm$ as \cite{Hollowood:2014rla,Sfetsos:2014lla}
\be
\label{currents.iso}
J_\pm=\l^{-1}A_\pm-A_\mp\,.
\ee 
Employing the above into \eqref{rsMaillet}, we find  \cite{Itsios:2014vfa}
\be
\label{rpmisolambda}
\begin{split}
&r_{\pm zw}\to r_{\pm zw}\,\Pi\,,\quad \Pi=\sum_A t_A\otimes t_A\,,\\
&r_{+zw}=-\frac{2\text{e}^2(1+z^2+x(1-z^2))zw}{(z-w)(1-z^2)}\,,\quad r_{-zw}=-r_{+wz}\,,
\\
&\text{where:}\quad x=\frac{1+\l^2}{2\l}\,,\quad \text{e}=\frac{2\l}{\sqrt{k(1-\l)(1+\l)^3}}\, ,
\end{split}
\ee
while the mYB equation \eqref{mYB} is also satisfied.
In addition, we note that the above expressions are invariant under the symmetry of the model
\be
\l\to\l^{-1}\,,\quad k\to-k\,.
\ee
The above result \eqref{rpmisolambda}, can be written in terms of a twist function $\varphi_\l(z)$ \cite{Hollowood:2015dpa}\footnote{
To get rid of the unphysical pole at $z=0$, we have transformed $z\to z^{-1}$.\label{zinverse}}
\be
\label{skfufsiusk}
r_{+zw}=-\frac{2}{z^{-1}-w^{-1}}\varphi^{-1}_\l(z^{-1})\,,
\ee
where
\be
 \varphi^{-1}_\l(z^{-1})=-\text{e}^2\,\frac{1+x+z^2(1-x)}{1-z^2}\,.
\ee
The twist function is a meromorphic one having poles of order one at 
\be
\varphi^{-1}_\l(z^{-1})=0\quad \Longrightarrow\quad  z_\pm=\pm\frac{1+\l}{1-\l}\,,
\ee
provided that $\l\neq0$, otherwise the expressions trivialize to zero and the non-ultra local term $\d_{12}'$ is removed \cite{Faddeev:1985qu}.
Evaluating the spatial Lax matrix \eqref{jrkdifyj}  and $s_{zw}$ at the poles we obtain
\be
L_1(z_\pm)=\pm J_\pm\,,\quad s_{z_\pm z_\pm}=\mp\frac1k\,,\quad s_{z_\pm z_\mp}=0\,.
\ee
Using the above in the Maillet brackets \eqref{rsMaillet}, we readily obtain the current algebra \eqref{KM}.\footnote{The $r_{zw}$ terms do not 
contribute as either they identically vanish or the commutators in \eqref{rsMaillet} vanish when 
they are evaluated at different poles or equal poles, respectively.} 

Next, we focus on two interesting zoom-in limits of the $\l$-deformed model, namely the non-Abelian T-dual of the PCM 
and the pseudo-dual chiral models. 
In the non-Abelian T-dual model, we expand $\l$ around one\cite{Sfetsos:2013wia}
\be
\l=1-\frac{\k^2}{k}\,,\quad k\gg1\,,
\ee
yielding a twist function
\be
\label{twpcm}
\varphi_\k^{-1}(z^{-1})=-\frac{1}{\k^2}\frac{1}{1-z^2}\,,
\ee
which coincides with the results
for the non-Abelian T-dual model in section 4.3.2 in \cite{Delduc:2019bcl}, by identifying $K=\kappa^2$ and transforming $z\to z^{-1}$.
This is also in agreement with the PCM result Eq.(35) in \cite{Maillet:1985ec}\footnote{This is expected since
the PCM model is canonically equivalent to its non-Abelian T-dual \cite{Curtright:1994be,Lozano:1995jx,Sfetsos:1996pm}.}, 
by identifying $\gamma=\nicefrac{2}{\k^2}$ and transforming $z\to z^{-1}$.  

\no
In the pseudodual chiral model, we expand $\l$ around minus one\cite{Georgiou:2016iom}
\be
\l=-1+\frac{b^{2/3}}{k^{1/3}}\,,\quad k\gg1\,,
\ee
yielding a twist function
\be
\varphi_b^{-1}(z^{-1})=-\frac{4}{b^2}\frac{z^2}{1-z^2}\, .
\ee
Note that, this twist function is basically the same as the one in \eqn{twpcm} for PCM as expected, since the pseudo-chiral and
PCM models have  their equations of motion and the Bianchi identities exchanged.

\subsection{The isotropic deformation at unequal levels}
\label{kdhksjsjjksn}

Let us consider $\l$-deformations of the direct product of two current
algebras at levels $k_1$ and $k_2$.
This model is described through \cite{Georgiou:2017jfi}
\begin{equation}
\label{wsksasjja}
\begin{split}
&S_{k_1,k_2;\l_1,\l_2}=S_{\text{WZW},k_1}(\frak{g}_1)+S_{\text{WZW},k_2}(\frak{g}_2)+\\
&+\frac{1}{\pi}\int_{\del B} \text{d}^2\sigma\, \Tr\left[\left(J_{1+}\phantom{0} J_{2+}\right) \binom{k_1\L_{21}\l_1D_2^T\l_2 \quad k_2\l_0\L_{21}\l_1}{k_1\l_0^{-1}\L_{12}\l_2\quad k_2\L_{12}\l_2D_1^T\l_1} \binom{J_{1-}}{J_{2-}}\right]\,,
\end{split}
\end{equation}
and  
\be
\L_{12}=(\mathbb{1}-\l_2D_1^T\l_1D_2^T)^{-1}\,,\quad \l_0=\sqrt{\frac{k_1}{k_2}}\,,
\ee
where
$J_{i\pm}=J_\pm(\frak{g}_i)\,, D_i=D(\frak{g}_i)$, with $\frak{g}_i\in G\,,\quad  i=1,2.$
 We choose without loss of generality that $k_2>k_1$, so that $\l_0<1$. 
For isotropic couplings, namely $\left(\lambda_{1,2}\right)_{AB}=\l_{1,2}\,\d_{AB}$, the equations of motion of \eqref{wsksasjja} take the form \cite{Georgiou:2017jfi}
\be
\del_\pm A_\mp=\pm\frac{1-\l^{\mp1}_0\l_1}{1-\l_1^2}[A_+,A_-]\,,\quad
\del_\pm B_\mp=\pm\frac{1-\l^{\pm1}_0\l_2}{1-\l_2^2}[B_+,B_-]\,,
\ee
where 
\begin{equation}
\begin{split}
&A_+ = i (\mathbb{1}-\l_1\l_2 D_1 D_2)^{-1}\l_1 ( \l_0 J_{1+} +  \l_2 D_1 J_{2+})\ ,
\\
&A_- = -i (\mathbb{1}-\l_1\l_2 D_2^T  D_1^T)^{-1}\l_1 ( \l_0^{-1} J_{2-} + \l_2D^T_2 J_{1-})\,,
\\
& B_+ = i (\mathbb{1}-\l_1\l_2 D_2 D_1)^{-1}\l_2 ( \l_0^{-1}J_{2+} + \l_1 D_2 J_{1+})\ ,
\\
&
B_- = -i (\mathbb{1}-\l_1\l_2 D_1^T D_2^T)^{-1} \l_2 (  \l_0 J_{1-} + \l_1 D^{T}_1 J_{2-})\ .
\end{split}
\end{equation}
The corresponding Lax matrices read
\be
L_\pm=\frac{2z}{z\mp1}\frac{1-\l_0^{\mp1}\l_1}{1-\l_1^2}A_\pm\,,\quad
\widehat L_\pm=\frac{2w}{w\mp1}\frac{1-\l_0^{\pm1}\l_2}{1-\l_2^2}B_\pm\,,
\quad (z,w)\in\mathbb{C}\, .
\ee
To evaluate \eqref{Sklyanin}, we need the Poisson brackets of $(A_\pm,B_\pm)$ which can be expressed in 
terms of two commuting copies of the current algebras
\be
\label{KM2}
\begin{split}
&\{J_{i\pm}^A(\s_1),J_{j\pm}^B(\s_2)\}_\text{P.B.}=\frac{2}{k_i}\d_{ij}\left(f_{ABC}J_{i\pm}^C(\s_2)\d_{12}\pm\d_{AB}\d'_{12}\right)\,,
\\
&\{J_{i\pm}^A(\s_1), J_{j\mp}^B(\s_2)\}_\text{P.B.}=0\,,
\end{split}
\ee
where the currents are given by \cite{Georgiou:2017jfi}
\be
\begin{split}
&J_{1+}=\l_0^{-1}\l_1^{-1}A_+-A_-\,,\quad J_{1-}=\l_0^{-1}\l_2^{-1}B_--B_+\,,\\
&J_{2+}=\l_0\l_2^{-1}B_+-B_-\,,\quad J_{2-}=\l_0\l_1^{-1}A_--A_+\,.
\end{split}
\ee
Employing the above into \eqref{rsMaillet} we find
\ba
\label{rpmanisolambda}
&&r_{\pm zw}\to r_{\pm zw}\,\Pi\,,\quad \hat r_{\pm zw}\to \hat r_{\pm zw}\,\Pi\,,\quad
\Pi=\sum_A t_A\otimes t_A\,,
\nonumber
\\
&&r_{+zw}=\frac{4\l_1zw(-2z\l_1+\l_0(z+1+(z-1)\l_1^2))(-2z\l_1+\l^{-1}_0(z-1+(z+1)\l_1^2))}{\sqrt{k_1k_2}(z-w)(1-z^2)(1-\l_1^2)^3}\,,
\nonumber
\\
&&\hat r_{+zw}=\frac{4\l_2zw(-2z\l_2+\l^{-1}_0(z+1+(z-1)\l_2^2))(-2z\l_2+\l_0(z-1+(z+1)\l_2^2))}{\sqrt{k_1k_2}(z-w)(1-z^2)(1-\l_2^2)^3}\,,
\nonumber
\\
&&r_{-zw}=-r_{+wz}\,,\quad \hat r_{-zw}=-\hat r_{+wz}\, .
\ea
We have checked that indeed the mYB equations \eqref{mYB} are satisfied.
Let us also note that \eqref{rpmanisolambda} is invariant under the symmetry \cite{Georgiou:2017jfi}
\be
\l_1\to\l_1^{-1}\,,\quad \l_2\to\l_2^{-1}\,,\quad k_1\to -k_2\,,\quad k_2\to-k_1\ 
\ee
and that it matches \eqref{rpmisolambda} for equal levels. This is so because as was shown in \cite{Georgiou:2017oly} the model with $k_1=k_2$, 
constructed in \cite{Georgiou:2016urf}, is canonically equivalent to two single $\l$-deformed models.

The above results could also be written in terms of two twist functions $(\varphi_\l,\widehat\varphi_\l)$ as in \eqref{skfufsiusk}, with
\be
\label{twistdifferent}
\begin{split}
&\varphi^{-1}_\l(z^{-1})=\frac{2\l_1(-2z\l_1+\l_0(z+1+(z-1)\l_1^2))(-2z\l_1+\l^{-1}_0(z-1+(z+1)\l_1^2))}{\sqrt{k_1k_2}(1-z^2)(1-\l_1^2)^3}\,,\\
&\widehat\varphi^{-1}_\l(z^{-1})=\frac{2\l_2(-2z\l_2+\l^{-1}_0(z+1+(z-1)\l_2^2))(-2z\l_2+\l_0(z-1+(z+1)\l_2^2))}{\sqrt{k_1k_2}(1-z^2)(1-\l_2^2)^3}\,.
\end{split}
\ee
These meromorphic twist functions have single poles at
\be
\label{twistdifferentzeros}
\begin{split}
&z_{+}=\frac{1-\l_1^2}{1-2\l_0\l_1+\l_1^2}\,,\quad z_{-}=-\frac{1-\l_1^2}{1-2\l^{-1}_0\l_1+\l_1^2}\,,
\\
& \hat z_{+}=\frac{1-\l_2^2}{1-2\l^{-1}_0\l_2+\l_2^2}\,,\quad \hat z_{-}=-\frac{1-\l_2^2}{1-2\l_0\l_2+\l_2^2}\,,
\end{split}
\ee
provided that $\l_{1,2}\neq0$, otherwise the expressions trivialize to zero and the 
non-ultra local term $\d_{12}'$ is removed \cite{Faddeev:1985qu}.
The poles in \eqref{twistdifferentzeros} are of order one, provided that $\l_{1,2}\neq\l_0,\l_0^{-1}$ otherwise they become of order two.
Assuming they are of order one, one can evaluate the spatial Lax matrices and $(s_{zw},\hat s_{zw})$ at the poles to obtain
\ba
&&L_1(z_+)=J_{1+}\,,\quad L_1(z_-)=-J_{2-}\,,\quad \widehat L_1(\hat z_+)=-J_{1-}\,,\quad \widehat L_1(\hat z_-)=J_{2+}\,,\\
&&s_{z_+z_+}=-\frac{1}{k_1}\,,\quad s_{z_-z_-}=\frac{1}{k_2}\,,\quad s_{z_\pm z_\mp}=0\,,\quad
\hat s_{\hat z_+\hat z_+}=\frac{1}{k_1}\,,\quad \hat s_{\hat z_-\hat z_-}=-\frac{1}{k_2}\,,\quad \hat s_{\hat z_\pm \hat z_\mp}=0\,.\nonumber
\ea
Using the above expressions in the Maillet brackets \eqref{rsMaillet}, we readily obtain the current algebra \eqref{KM2}.

\no
Let us now analyze  the particular case in which $\l_1=\l_0=\l_2$. 
At that point the twist functions have poles of order two at 
$z=1$ and $\hat z=-1$. Upon evaluating the spatial Lax matrices and $s_{zw},\hat s_{zw}$ at 
 these poles we obtain that
\be
\begin{split}
&L_1=\frac{1}{k_1-k_2}\left(k_1 J_{1+}+k_2 J_{2-}\right)\,,\quad \widehat L_1=-\frac{1}{k_1-k_2}\left(k_1 J_{1-}+k_2 J_{2+}\right)\,,\\
&s_{z z}=-\frac{1}{k_1-k_2}\,,\quad \hat s_{\hat z\hat z}=\frac{1}{k_1-k_2}\,.
\end{split}
\ee
Using the above in the Maillet brackets \eqref{rsMaillet}, we obtain two copies of a current algebra at level $k_2-k_1$
\be
\begin{split}
&\{L^A_1(\s_1),L^B_1(\s_2)\}_\text{P.B.}=\frac{2}{k_1-k_2}\left(f_{ABC}\,L_1^C(\s_2)\,\d_{12}+\d_{AB}\,\d'_{12}\right)\,,\\
&\{\widehat L^A_1(\s_1),\widehat L^B_1(\s_2)\}_\text{P.B.}=
\frac{2}{k_2-k_1}\left(f_{ABC}\,\widehat L_1^C(\s_2)\,\d_{12}+\d_{AB}\,\d'_{12}\right)\,.
\end{split}
\ee
These current algebras are realised at level $k_2\!-\! k_1$. This is in resonance with the fact that the fixed point in the IR
when $\l_1=\l_0=\l_2$ is given by the CFT  $\frac{G_{k_1}\times G_{k_2-k_1}}{G_{k_2}}\times G_{k_2-k_1}$ \cite{Georgiou:2017jfi}.

\section{Coset spaces}
\label{coset.strong.section}

In this section we first revisit, as a warm up, the construction of \cite{Hollowood:2015dpa} of the $r/s$-Maillet form for the symmetric coset $G_k/H_k$ case. 
Then we proceed to find the same form for the coset $G_{k_1}\times G_{k_2}/G_{k_1+k_2}$.

\subsection{The isotropic $G_k/H_k$ symmetric coset space}
\label{iso.symmetric.coset}

Let us consider a semi-simple group $G$ and its decomposition to a semi-simple group $H$ and a symmetric coset $G/H$. 
Consider now the action \eqref{klfjslkskdjdw} for a coupling 
matrix $\l_{AB}$ with elements $\l_{ab}=\d_{ab}$ and $\l_{\a\b}=\l\, \d_{\a\b}$, which is known to be integrable \cite{Hollowood:2014rla}. The subgroup indices are denoted by Latin letters and coset indices by Greek letters. 
The restriction to symmetric cosets amounts to structure constants with $f_{\a\b\g}=0$. The equations of motion of \eqref{klfjslkskdjdw}, simplify
to \cite{Hollowood:2014rla}
\be
\label{eomGH}
\begin{split}
&\del_+ A_--\del_-A_+=[A_+,A_-]+\l^{-1}[B_+,B_-]\,,\\
&\del_\pm B_\mp=-[B_\mp,A_\pm]\,,\\
&\text{where:}\quad A_\pm=A^a_\pm t_a\,,\quad B_\pm=B^\a_\pm t_\a\,,
\end{split}
\ee
and the corresponding Lax matrices reads 
\be
L_\pm=A_\pm+\frac{z^{\pm1}}{\sqrt{\l}}B_\pm\,,\quad z\in\mathbb{C}\,.
\ee
Using the above we construct the spatial Lax matrix
\be
\label{spatialsymmetric}
L_1=A_+-A_-+\frac{1}{\sqrt{\l}}\left(z B_+-z^{-1} B_-\right)\,.
\ee
To evaluate \eqref{Sklyanin}, we split the algebra \eqref{KM} into subgroup and coset components
\be
\label{KMsymmetric}
\begin{split}
&\{j_\pm^a(\s_1),j_\pm^b(\s_2)\}_\text{P.B.}=\frac{2}{k}\left(f_{abc}\,j^c_\pm(\s_1)\,\d_{12}\pm\d_{ab}\,\d_{12}'\right)\,,\\
&\{J_\pm^\a(\s_1),J_\pm^\b(\s_2)\}_\text{P.B.}=\frac{2}{k}\left(f_{\a\b c}\,j^c_\pm(\s_1)\,\d_{12}\pm\d_{\a\b}\,\d_{12}'\right)\,,\\
&\{j_\pm^a(\s_1),J_\pm^\b(\s_2)\}_\text{P.B.}=\frac{2}{k} f_{a\b\g}J^\g_\pm(\s_1)\d_{12}\,,
\end{split}
\ee
where the currents 
\be
j_\pm=A_\pm-A_\mp\,,\quad J_\pm=\l^{-1}B_\pm-B_\mp\,.
\ee
Note that there is the first class constraint
\be
\label{first.symmetric}
\chi=j_++j_-\approx0\, .
\ee
Since $\{\chi^a(\s_1),\chi^b(\s_2)\}_\text{P.B.}\approx0$ this corresponds to the residual gauge symmetry.
Before we proceed to the computation of the Poisson brackets of \eqref{spatialsymmetric}
we could also include $\chi$ into the spatial Lax matrix as\footnote{This idea was employed in \cite{Magro:2008dv} and further developed in \cite{Vicedo:2009sn},  in the context of the classical exchange algebra for the pure-spinor superstring on $\text{AdS}_5 \times S^5$ \cite{Mikhailov:2007eg}.}
\be
\label{spatialsymmetricN}
\widetilde L_1 =L_1+\varrho(\l,z)\,\chi\,,
\ee
where $\varrho$ is an arbitrary, Lagrange multiplier type, function of $\l$ and $z$ to be later determined.
We can now compute the Poisson brackets of $\widetilde L_1$ and after some tedious algebra we obtain
the {\it weak} analogue of the $r/s$ Maillet form \eqref{rsMaillet}, namely that
\be
\label{weakMaillet}
\{\widetilde L_1^{(1)}(\sigma_1;z),\widetilde L_1^{(2)}(\sigma_2;w)\}_\text{P.B.}\approx
\left([r_{-zw},\widetilde L_1^{(1)}(\s_1;z)]+[r_{+zw},\widetilde L_1^{(2)}(\s_2;w)]\right)\d_{12}-2s_{zw}\d'_{12}\,,
\ee
where
\be
\label{rssymmetric}
\begin{split}
&r_{\pm zw}\to r^\text{sub}_{\pm zw}\Pi_\text{sub}+r^\text{coset}_{\pm zw} \Pi_\text{coset}\,,\quad
\Pi_\text{sub}=\sum_a t_a\otimes t_a\,,\quad \Pi_\text{coset}=\sum_\a t_\a\otimes t_\a\,,\\
&r^\text{sub}_{+ zw}=\frac{2(z^2-\l^{-1})(w^2-\l)}{k(z^2-w^2)(\l-\l^{-1})}-\frac{2\varrho(\l,z)}{k}\,,\quad 
r^\text{coset}_{+ zw}=\frac{2w(z^2-\l^{-1})(z^2-\l)}{kz(z^2-w^2)(\l-\l^{-1})}\,,\\
&r^\text{sub}_{-zw}=-r^\text{sub}_{+wz}\,,\quad r^\text{coset}_{-zw}=-r^\text{coset}_{+wz}\,.
\end{split}
\ee
The above expressions are not apparently invariant under the symmetry of the model
\be
\label{sduality}
\l\to\l^{-1}\,,\quad k\to-k\,,
\ee
as they transform as
\begin{equation}
r^\text{sub}_{\pm zw}\to r^\text{sub}_{\pm zw}\pm2\frac{1+\varrho(\l^{-1},z)+\varrho(\l,z)}{k}\,,\quad r^\text{coset}_{\pm zw}\to r^\text{coset}_{\pm zw}\,.
\end{equation}
Let us now check if the modified Yang--Baxter equation \eqref{mYB} is satisfied for \eqref{rssymmetric}, yielding that
\begin{equation}
\label{mYBgauge}
\begin{split}
&[r_{+z_1z_3}^{(13)},r_{-z_1z_2}^{(12)}]+[r_{+z_2z_3}^{(23)},r_{+z_1z_2}^{(12)}]+[r_{+z_2z_3}^{(23)},r_{+z_1z_3}^{(13)}]
\\
&\qq\quad =c_1(z_{1,2})\, f_{abc}t_a\otimes t_b \otimes t_c+c_2(z_{1,2})\,   f_{a\b\g} t_\b \otimes t_\g\otimes t_a\, ,
\end{split}
\ee
where 
\ba
&&c_1(z_{1,2})=\frac{2}{k}\left(-\varrho(\l,z_2) r^\text{sub}_{+z_1z_2}+\varrho(\l,z_1)\left(r^\text{sub}_{-z_1z_2}-\frac2k\varrho(\l,z_2)\right)\right)\,,\\
&&c_2(z_{1,2})=\frac2k\left(-\varrho(\l,z_2)r^\text{coset}_{+z_1z_2}+\varrho(\l,z_1)r^\text{coset}_{-z_1z_2}\right)-\frac{4\l(z_1-\l^{-1}z^{-1}_1)(z_2-\l^{-1}z^{-1}_2)}{k^2(\l-\l^{-1})^2}\,.\nonumber
\ea
The non-vanishing 
right hand side \eqref{mYBgauge} can be attributed to the residual gauge symm\-etry of the subgroup. A similar situation was
encountered in \cite{Mikhailov:2007eg} where the Maillet algebra was computed in the presence of Wilson lines among
the inserted operators. 
The presence of the Wilson lines leads to non-trivial monodromy properties and the Jacobi identity is weakly satisfied.
Another similar situation has appeared before in studies of parafermions
 \cite{Bardakci:1990ad,Bardacki:1990wj}.\footnote{We would like to thank Sakura Sch\"afer-Nameki for discussions on this point.} 

\no
In order, to proceed we demand that the mYB equation is satisfied, i.e. $c_{1,2}(z_{1,2})=0$. 
These two conditions determine $\varrho(\l,z_1)$ and $\varrho(\l,z_2)$. It turns out that they both originate from
the following two solutions
\be
\label{fujjddksk}
\varrho_1(\l,z)=\frac{\l^{-1}-z^2}{\l-\l^{-1}}\quad\text{or}\quad \varrho_2(\l,z)=\frac{z^{-2}-\l}{\l-\l^{-1}}\, ,
\ee
which satisfy $\varrho_1(\l^{-1},z^{-1})=\varrho_2(\l,z)$, hence we analyze the above results only for $\varrho_1$.
 Then \eqref{rssymmetric}  simplifies to
\be
r^\text{sub}_{+ zw}=\frac{2\l(z^2-\l^{-1})(z^2-\l)}{k(z^2-w^2)(\l-\l^{-1})}\,,\quad 
r^\text{coset}_{+ zw}=\frac{2w(z^2-\l^{-1})(z^2-\l)}{kz(z^2-w^2)(\l-\l^{-1})}\,,
\ee
which is now invariant under the symmetry \eqref{sduality}.

\no
This result admits a twist function description\cite{Vicedo:2010qd}
\be
\label{jfkshfhdj}
r^\text{sub}_{+zw}=-\frac{2z^2}{z^2-w^2}\varphi_\l^{-1}(z)\,,\quad r^\text{coset}_{+zw}=-\frac{2zw}{z^2-w^2}\varphi_\l^{-1}(z)\,,
\ee
where $\varphi_\l(z)$ is given by \cite{Hollowood:2015dpa}
\be
\label{dusjfnd}
\varphi_\l(z)=-\frac{kz^2(\l-\l^{-1})}{(z^2-\l)(z^2-\l^{-1})}\,,
\ee
with poles of order one at
\be
z_{1\pm}=\pm\l^{-1/2}\,,\quad z_{2\pm}=\pm\l^{1/2}\,,
\ee
provided that $\l\neq0$.
A comment is in order regarding the twist function description \eqref{jfkshfhdj}.
Demanding that \eqref{rssymmetric} takes the form of \eqref{jfkshfhdj}, fixes $\varrho$
and $\varphi_\l(z)$ to the value $\varrho_1$ in \eqref{fujjddksk} and \eqref{dusjfnd}, respectively.

\no
Evaluating the spatial Lax matrix \eqref{spatialsymmetricN} and $s_{zw}$  at these poles we obtain that
\begin{equation}
\begin{split}
&\widetilde L_1(z_{1\pm})=j_+\pm J_+\,,\quad \widetilde L_1(z_{2\pm})=-j_-\mp J_-\,,\\
& s^\text{sub}_{z_{1\pm}z_{1\pm}}=-\frac1k\,,\quad s^\text{sub}_{z_{2\pm}z_{2\pm}}=\frac1k\,,\quad
s^\text{sub}_{z_{1\pm}z_{1\mp}}=-\frac1k\,,\quad s^\text{sub}_{z_{2\pm}z_{2\mp}}=\frac1k\,,\\
& s^\text{coset}_{z_{1\pm}z_{1\pm}}=-\frac1k\,,\quad s^\text{coset}_{z_{2\pm}z_{2\pm}}=\frac1k\,,\quad
s^\text{coset}_{z_{1\pm}z_{1\mp}}=\frac1k\,,\quad s^\text{coset}_{z_{2\pm}z_{2\mp}}=-\frac1k\,,\\
&s^\text{sub}_{z_{1\pm}z_{2\pm}}=0=s^\text{sub}_{z_{1\pm}z_{2\mp}}\,,\quad 
s^\text{coset}_{z_{1\pm}z_{2\pm}}=0=s^\text{coset}_{z_{1\pm}z_{2\mp}}\,.
\end{split}
\end{equation}
Plugging the above in the Maillet brackets \eqref{weakMaillet}, we uncover the current algebra \eqref{KMsymmetric}.
Let us mention that for $\l=0$, these expressions drastically simplify to
\be
\widetilde L_1=-j_-\,,\quad r_{+zw}^\text{sub}=\frac{2z^2}{k(z^2-w^2)}\,,\quad r_{+zw}^\text{coset}=\frac{2zw}{k(z^2-w^2)}\,,\quad
s_{zw}^\text{sub}=\frac1k\,,\quad s_{zw}^\text{coset}=0\ 
\ee
and the non-ultralocal term is partially removed in the subgroup part of the algebra \cite{Delduc:2012qb,Hollowood:2015dpa}.

\subsection{The isotropic $G_{k_1}\times G_{k_2}/G_{k_1+k_2}$ coset space}
\label{diagonal.symmetric.coset}

As a second example we consider the isotropic $\l$-deformed $G_{k_1}\times G_{k_2}/G_{k_1+k_2}$ coset space.
This has been proved to be integrable as it admits a flat Lax connection \cite{Sfetsos:2017sep}. 
The action of this model reads \cite{Sfetsos:2017sep}
\begin{equation}
\begin{split}
& S_{k,\l}= S_{\text{WZW},k_1}(\frak{g}_1) + S_{\text{WZW},k_2}(\frak{g}_2)
\\
& \quad
+ {k\ov \pi} \int_{\del B} \text{d}^2\s\,\text{Tr} \Big[  s_1\, J_{1+} \L_{12}^{-T} \left((1-\l)(s_1J_{1-}+ s_2\,J_{2-})- 4 s_1s_2 \l(D_2^T-\mathbb{1})J_{1-}\right)
\\
&
\quad\quad\quad\quad + s_2\, J_{2+}\, \L_{21}^{-T} \left((1-\l)(s_1J_{1-}+ s_2J_{2-})-
4 s_1s_2 \l(D_1^T-\mathbb{1})J_{2-}\right) \Big] \ .
\end{split}
\end{equation}
with
\be
\begin{split}
&\L_{12}=4 \l s_1 s_2 (D_1-\mathbb{1})(D_2-\mathbb{1}) + (\l-1)\left(s_1D_1+s_2D_2-\mathbb{1})\right)\,,\\
& s_i=\frac{k_i}{k}\,,\quad k=k_1+k_2\,,\quad i=1,2\,.
\end{split}
\ee
Its equations of motion take the form of
\be
\label{eomGGovG}
\begin{split}
&\del_+\widehat{\cal A}_--\del_+\widehat{\cal A}_-=[\widehat{\cal A}_+,\widehat{\cal A}_-]+(\a^2+\b)[{\cal B}_+,{\cal B}_-]\,,\\
&\del_\pm{\cal B}_\mp=-[{\cal B}_\mp,\widehat {\cal A}_\pm]\,,
\end{split}
\ee
where
\be
\widehat {\cal A}_\pm={\cal A}_\pm+\a {\cal B}_\pm\,,\quad
{\cal A}_\pm=\frac12\left(A_{1\pm}+A_{2\pm}\right)\,,\quad {\cal B}_\pm=\frac12\left(A_{1\pm}-A_{2\pm}\right)\,,
\ee
with 
\be
\a=-\frac{(s_1-s_2)(1-\l)}{1-\l_f^{-1}\l}\,,\quad\beta=\frac{1+\l-2(1-4s_1s_2)\l^2}{\l(1-\l_f^{-1}\l)}\,,\quad
\l_f^{-1}=1-8s_1s_2\,.
\ee
and
\be
\begin{split}
& A_{1+}=i \L_{21}^{-1}\left((1-\l)(s_1\,J_{1+}+ s_2\,J_{2+})- 4 s_1s_2 \l(D_2-\mathbb{1})J_{1+}\right)\,,
\\
&
A_{2+}=i \L_{12}^{-1}\left((1-\l)(s_1\,J_{1+}+ s_2\,J_{2+})- 4 s_1s_2 \l(D_1-\mathbb{1})J_{2+}\right)\,,
\\
& A_{1-}= - i \L_{12}^{-T}\left((1-\l)(s_1\,J_{1-}+ s_2\,J_{2-})- 4 s_1s_2 \l(D_2^T-\mathbb{1})J_{1-}\right)\,,
\\
&
A_{2-}= - i \L_{21}^{-T}\left((1-\l)(s_1\,J_{1-}+ s_2\,\,J_{2-})- 4 s_1s_2 \l(D_1^T-\mathbb{1})J_{2-}\right)\,.
\end{split}
\ee
The system of equations \eqref{eomGGovG} admits a flat Lax connection \cite{Sfetsos:2017sep}
\be
L_\pm=\widehat {\cal A}_\pm+z^{\pm1}\sqrt{\a^2+\b}\,{\cal B}_\pm= {\cal A}_\pm+\left(\a+z^{\pm1}\sqrt{\a^2+\b}\right)\,{\cal B}_\pm\,,\quad z\in\mathbb{C}\,.
\ee
Using the above we build the spatial Lax matrix
\be
\label{spatialdiagonal}
L_1=\frac12\left((1+\a)A_{1|1}+(1-\a)A_{2|1}\right)+\sqrt{\a^2+\b}\left(z{\cal B}_+-z^{-1}{\cal B}_-\right)\,,
\ee
where $A_{i|1}=A_{i+}-A_{i-}$.
To evaluate \eqref{Sklyanin},  we need the Poisson brackets of $A_{i\pm}$ which in turn can be obtained from
 two commuting copies of current algebras \eqref{KM2}, where \cite{Sfetsos:2017sep}
\be
\label{Jsdiagonal}
\begin{split}
&J_{1+}=\frac{1}{2s_1}(\l^{-1}-1){\cal B}_++A_{1|1}\,,\quad J_{1-}=\frac{1}{2s_1}(\l^{-1}-1){\cal B}_+-A_{1|1}\,,\\
&J_{2+}=-\frac{1}{2s_2}(\l^{-1}-1){\cal B}_++A_{2|1}\,,\quad J_{2-}=-\frac{1}{2s_2}(\l^{-1}-1){\cal B}_+-A_{2|1}\,.
\end{split}
\ee
To proceed we define the subgroup generators ${\cal H}_\pm$
\be
\label{kflsirms}
{\cal H}_\pm=\frac12\left(s_1 J_{1\pm}+s_2 J_{2\pm}\right)\,,
\ee
or equivalently through \eqref{Jsdiagonal} 
\be
\label{jkdkslsl}
{\cal H}_\pm=\pm\frac12\left(s_1 A_{1|1}+s_2 A_{2|1}\right)\,.
\ee
Similarly to \eqref{first.symmetric}, we find a first class constraint
\be
\chi={\cal H}_++{\cal H}_-\approx0\, .
\ee
Since $\{\chi^A(\s_1),\chi^B(\s_2)\}_\text{P.B.}\approx0$, we have again a residual gauge symmetry.
Before we proceed to the derivation of the Poisson brackets of \eqref{spatialdiagonal},
we include $\chi$ into the spatial Lax matrix as in \eqref{spatialsymmetricN}
\be
\label{spatialdiagonalN}
\widetilde L_1 =L_1+\varrho(\l,z)\,\chi\,.
\ee
Then we express $\widetilde L_1$ or equivalently $A_{i|1},{\cal B}_\pm$ in terms of $J_{i\pm}$.\footnote{
We first rewrite ${\cal H}_+$ in \eqref{jkdkslsl} as
\begin{equation*}
\label{fjksxudjn}
{\cal H}_+=\frac12A_{1|1}-s_2\left({\cal B}_+-{\cal B}_-\right)\,,
\end{equation*}
and we also note that \eqref{jkdkslsl} can be also used to express $A_{2|1}$ in terms of $A_{1|1}$ as
\begin{equation*}
\label{yfuyekmdydh}
A_{2|1}=\frac{1}{s_2}\left(2{\cal H}_+-s_1 A_{1|1}\right)\,.
\end{equation*}
Using the above, \eqref{kflsirms} and $J_{1\pm}$'s in \eqref{Jsdiagonal}, we can write $({\cal B}_\pm,A_{1|1})$ in terms of $(J_{1\pm},J_{2+})$.
Therefore, the $\widetilde L_1$ is expressed on the $J_{i\pm}$ basis.
}
We are in position to compute the Poisson brackets of $\widetilde L_1$ and for simplicity we start with the equal level case.

\subsubsection*{The equal level case}

Plugging the spatial Lax matrix \eqref{spatialdiagonalN} into the Maillet algebra \eqref{weakMaillet}, we find
\be
\label{requaldiagonal}
\begin{split}
&r_{\pm zw}\to r_{\pm zw}\Pi\,,\quad \Pi=\sum_A t_A\otimes t_A\,,\\
&r_{+zw}=\frac1k\left(\frac{2(z-\sqrt{\l})(z^2-\l^{-1})(w+\sqrt{\l})}{z(z-w)(\l-\l^{-1})}-\varrho(\l,z)\right)
\,,\quad r_{-zw}=-r_{+wz}\,.
\end{split}
\ee
The above expression transforms under the symmetry \eqref{sduality}, as
\be
\label{symmetry.diagonal.iso.coset}
r_{+zw}\to r_{+zw}+\frac{2-\frac{2\sqrt{\l}}{1+\l}(z+z^{-1})+\varrho(\l^{-1},z)+\varrho(\l,z)}{k}\,.
\ee
Then we plug  \eqref{requaldiagonal} into the mYB equation \eqref{mYB}, yielding
\be
\label{mYBgaugeN}
\begin{split}
&[r_{+z_1z_3}^{(13)},r_{-z_1z_2}^{(12)}]+[r_{+z_2z_3}^{(23)},r_{+z_1z_2}^{(12)}]+[r_{+z_2z_3}^{(23)},r_{+z_1z_3}^{(13)}]=c(z_{1,2})\, f_{ABC}\,t_A\otimes t_B\otimes t_C\,,\\
&c(z_{1,2})=\frac1k\left(\varrho(\l,z_1) r_{-z_1z_2}-\varrho(\l,z_2) r_{+z_1z_2}\right)-\frac{1}{k^2}\varrho(\l,z_1)\varrho(\l,z_2)\,.
\end{split}
\ee
Let us now demand the right hand side of \eqref{mYBgaugeN} vanishes for every $z_{1,2}$, i.e. $c(z_{1,2})=0$. 
Demanding that this condition is satisfied in  the limit  $z_2\to z_1=z$, yields a first order differential equation for $\varrho(\l,z)$ with respect to $z$ 
which has the following solution
\be
\label{fhdidfksk}
\varrho(\l,z)=-\frac{2(z-\sqrt{\l})(z^2-\l^{-1})}{z(\l-\l^{-1}-2\l^{-1}(z+\sqrt{\l})f(\l))}\,,
\ee
where $f(\l)$ is a $\l$-dependent integration constant. Plugging the above into $c(z_{1,2})$ in \eqref{mYBgaugeN}, we find that it is satisfied for every $z_{1,2}$.
Then we use \eqref{fhdidfksk} and we find that \eqref{requaldiagonal} is invariant under the symmetry \eqref{sduality}, provided that
\be
f(\l)=\l^{-1/2}(1+\l^{-1})\frac{\l^4-h(\l)}{2\l^2}\,,\quad\text{with}\quad h(\l) h(\l^{-1})=1\,.
\ee
For simplicity we pick up the integration constant to be $f(\l)=0$, a choice which is also justified by the 
results that follow. This choice leads to
\be
\label{fhdidfksk1}
\varrho(\l,z)=-\frac{2(z-\sqrt{\l})(z^2-\l^{-1})}{z(\l-\l^{-1})}\ 
\ee
and
\be
r_{+zw}=\frac{2(z^2-\l)(z^2-\l^{-1})}{kz(z-w)(\l-\l^{-1})}\,,\quad r_{-zw}=-r_{+wz}\,.
\ee
The above expression can be written in terms of a twist function $\varphi_\l(z)$ \cite{Delduc:2015xdm}
\be
\label{rptwistdiagonal}
r_{+zw}=-\frac{1}{z-w}\varphi_\l(z)^{-1}\,,
\ee
where 
\be
\label{fjksfdk}
\varphi_\l(z)=-\frac{kz(\l-\l^{-1})}{2(z^2-\l^{-1})(z^2-\l)}\,.
\ee
The meromorphic twist function \eqref{fjksfdk} has four poles of order one at
\be
\label{YBzeros}
z_{1\pm}=\pm\l^{-1/2}\,,\quad z_{2\pm}=\pm\l^{1/2}\,,
\ee
provided that $\l\neq0$.
Evaluating the spatial Lax matrix \eqref{spatialdiagonalN} and $s_{zw}$ at \eqref{YBzeros}, using also \eqref{fhdidfksk1},  we obtain
\be
\begin{split}
&\widetilde L_1(z_{1+})=J_{1+}\,,\quad \widetilde L_1(z_{1-})=J_{2+}\,,\quad 
\widetilde L_1(z_{2+})=-J_{1-}\,,\quad \widetilde L_1(z_{2-})=-J_{2-}\,,\\
&s_{z_{1\pm}z_{1\pm}}=-\frac2k\,,\quad s_{z_{2\pm}z_{2\pm}}=\frac2k\,,\quad 
s_{z_{1\pm}z_{1\mp}}=0=s_{z_{2\pm}z_{2\mp}}\,,\quad 
s_{z_{1\pm}z_{2\pm}}=0=s_{z_{1\pm}z_{2\mp}}\,.
\end{split}
\ee
Employing the above in the Maillet brackets \eqref{weakMaillet}, we obtain the current algebra \eqref{KM2}.
At $\l=0$, the expressions drastically simplify respectively to
\be
\widetilde L_1=-\frac12\left(J_{1-}+J_{2-}\right)\,,\quad r_{+zw}=\frac{2z}{z-w}s_{zw}\,,\quad s_{zw}=\frac{1}{k}\,,
\ee
so the non-ultralocal term $\d'_{12}$ is partially removed \cite{Delduc:2012qb}.

Finally, two comments are in order regarding the twist function description \eqref{rptwistdiagonal}. 
Firstly, demanding that \eqref{requaldiagonal} takes the form \eqref{rptwistdiagonal}, uniquely fixes $\varrho$ and $\varphi_\l(z)$ to
the values \eqref{fhdidfksk1} and  \eqref{fjksfdk} respectively. This property will be the guideline in the unequal level case which follows. 
Secondly,  it was argued in  \cite{Delduc:2015xdm}, using the results of \cite{Delduc:2013fga}, that the one-parameter deformation
of the coset $\s$-model $G\times G/G_\text{diag}$ admits such a description in terms of  the twist function (see Eq.(3.3) of \cite{Delduc:2015xdm})
\be
\label{twistYB}
\varphi_\text{YB}(z)=\frac{16Kz}{(1-z^2)^2+\eta^2(1+z^2)^2}\,,
\ee
which after the analytic continuation
\begin{equation}
\label{fskddjkkfds}
\eta=i\frac{1-\l}{1+\l}\,,\quad \frac{8i K}{\eta}=k\,,
\end{equation}
\eqref{twistYB} maps to \eqref{fjksfdk}.
Such a mapping is expected since the $\l$ and the $\eta$ deformations are related up to a Poisson--Lie T-duality and an analytic continuation, as the one in 
\eqn{fskddjkkfds} (see \cite{Vicedo:2015pna,Hoare:2015gda,Sfetsos:2015nya,Klimcik:2015gba,Klimcik:2016rov}).

\subsubsection*{The unequal level case}

Upon inserting the spatial Lax matrix \eqref{spatialdiagonalN} into the Maillet algebra \eqref{weakMaillet}, we require
that $r_{\pm zw}$ can be expressed in terms of twist function description as in \eqref{rptwistdiagonal}, which automatically satisfies the mYB \eqref{mYB}. 
This is an arduous task whose end result takes the form 
\be
\begin{split}
\label{fisols}
&\varphi_\l(z)=
\frac{1}{\zeta^2}\frac{8ks_1s_2\xi z}{z^4+1+4\xi^2\frac{s_1^2-s_2^2}{\zeta}z(z^2+1)+\left(2+4\frac{(s_1^2-s_2^2)^2\xi^4-1}{\zeta^2}\right)z^2}\,,\\
&\zeta=8s_1s_2\frac{\sqrt{\l(1-\l_1^{-1}\l)(1-\l_2^{-1}\l)(1-\l_3^{-1}\l)}}{(1-\l_f^{-1}\l)^2}\,,\quad \xi=\frac{1-\l}{1-\l_f^{-1}\l}\,,\\
&\l_1=\frac{1}{s_2-3s_1}\,,\quad \l_2=\frac{1}{s_1-3s_2}\,,\quad\l_3=\frac{1}{(s_1-s_2)^2}\,,\quad \l_f^{-1}=1-8s_1s_2\,.
\end{split}
\ee
For equal levels the above expressions are in agreement with \eqref{fjksfdk}. The twist function \eqref{fisols} is invariant under the remarkable transformation \cite{Sfetsos:2017sep}
\be
\l\to\frac{1-(s_1-s_2)^2\l}{(s_1-s_2)^2-\l_f^{-1}\l}\,,\quad k_1\to-k_1\,,\quad k_2\to-k_2\,,
\ee
as one can readily check using the induced transformations $(\xi,\zeta)\to(-\xi,\zeta)$.

The meromorphic twist function \eqref{fisols} has four poles of order one at
\be
\label{fhsskfhfdjd}
z_{1\pm}=\pm\zeta^{-1}(1\mp(s_1-s_2)\xi)(1+\xi)\,,\quad 
z_{2\pm}=\pm\zeta^{-1}(1\pm(s_1-s_2)\xi)(1-\xi)\,,
\ee
provided that $\zeta\neq0$ or equivalently that $\l$ is different from the CFT points \cite{Sfetsos:2017sep}
\be
\label{xcjosso}
\l\neq\left\{0,\l_1,\l_2,\l_3\right\}\,.
\ee 
Evaluating the spatial Lax matrix \eqref{spatialdiagonalN} and $s_{zw}$  at the poles \eqref{fhsskfhfdjd}, we obtain
\be
\begin{split}
&\widetilde L_1(z_{1+})=J_{1+}\,,\quad \widetilde L_1(z_{1-})=J_{2+}\,,\quad 
\widetilde L_1(z_{2+})=-J_{1-}\,,\quad \widetilde L_1(z_{2-})=-J_{2-}\,,\\
&s_{z_{1+}z_{1+}}=-\frac{1}{k_1}\,\quad s_{z_{1-}z_{1-}}=-\frac{1}{k_2}\,,\quad
s_{z_{2+}z_{2+}}=\frac{1}{k_1}\,\quad s_{z_{2-}z_{2-}}=\frac{1}{k_2}\,,\\
&s_{z_{1\pm}z_{1\mp}}=0=s_{z_{2\pm}z_{2\mp}}\,,\quad 
s_{z_{1\pm}z_{2\pm}}=0=s_{z_{1\pm}z_{2\mp}}\,.
\end{split}
\ee
Using the above in the Maillet brackets \eqref{weakMaillet}, we readily obtain the current algebra \eqref{KM2}.
At the CFT points \eqref{xcjosso}, the expressions for the spatial Lax and $r_{zw}$ drastically simplify to
\be
\begin{split}
&\widetilde L_1=\left\{-s_1J_{1-}-s_2J_{2-}\,,\frac{s_1J_{1+}+s_2J_{2-}}{s_1-s_2}\,,-\frac{s_1J_{1-}+s_2J_{2+}}{s_1-s_2}\,,s_1J_{1+}+s_2J_{2+}\right\}\,,\\
&r_{+zw}=\frac{2z}{z-w}s_{zw}\,,\quad
s_{zw}=\left\{\frac1k,\frac{1}{k_2-k_1}\,,\frac{1}{k_1-k_2}\,,-\frac1k\right\}\, ,\quad  k= k_1+k_2\, ,
\end{split}
\ee
in the ordering of CFT points given by \eqn{xcjosso}.
Therefore,  the non-ultralocal term $\d'_{12}$ is partially removed at the CFT points \eqref{xcjosso}.

Finally, three comments are in order regarding the twist function description \eqref{fisols}:
\begin{enumerate}
\item
The above twist function \eqref{fisols} matches the twist function for the two parameter deformation of $G\times G/G_\text{diag}$, or equivalently the bi-Yang--Baxter model, see Eq.(3.8) of \cite{Delduc:2015xdm}
\be
\label{twistbiYB}
\begin{split}
&\varphi_\text{bi-YB}(z)=
\frac{1}{\zeta^2}\frac{16Kz}{z^4+1+\frac{\eta^2-\tilde\eta^2}{\zeta}z(z^2+1)+\left(2+\frac{(\eta^2-\tilde\eta^2)^2-16}{4\zeta^2}\right)z^2}\,,\\
&\zeta=\sqrt{\left(1+\frac{1}{4}\left(\eta-\tilde\eta\right)^2\right)\left(1+\frac{1}{4}\left(\eta+\tilde\eta\right)^2\right)}\,,
\end{split}
\ee
up to the analytic continuation 
\be
\label{kffhjsks}
\eta=2is_2\xi\,,\quad \tilde\eta=2is_1\xi\,,\quad K=\frac12ks_1s_2\xi\,.
\ee
Agreement with the one parameter deformation of the coset $\s$-model  $G\times G/G_\text{diag}$ of Eqs.\eqref{twistYB}, \eqref{fskddjkkfds} is found for $\tilde\eta=\eta$.
\item
Using the results of \cite{Delduc:2013fga,Delduc:2015xdm} along with \eqref{kffhjsks}, the monodromy matrix when evaluated at the poles $z_{i\pm}$ generates a $q$-deformed Poisson algebra with
\be
q_1=\text{e}^{-i/{k_1}}\,,\quad q_2=\text{e}^{-i/{k_2}}\,,
\ee
being roots of unity \cite{Hollowood:2014rla}. 
\item
Using Eq.(4.9) of \cite{Sfetsos:2017sep} and \eqref{kffhjsks}, we can find the RG flows for two-parameter deformed $G\times G/G_\text{diag}$
\be
\label{RG.eta.diagonal}
\begin{split}
& \frac{\text{d}\eta}{\text{d}t}=\frac{c_G\,\eta}{32K}\left(1+\frac14\left(\eta+\tilde\eta\right)^2\right)\left(1+\frac14\left(\eta-\tilde\eta\right)^2\right)\,,
\\
& 
\frac{\text{d}}{\text{d}t}\left(\frac{\eta}{\tilde\eta}\right)=0\,,\quad
\frac{\text{d}}{\text{d}t}\left(\frac{\eta}{K}\right)=0\,,
\end{split}
\ee
where $t=\ln\mu^2$ is the logarithmic RG scale. Please note that \eqref{RG.eta.diagonal} is in agreement  with the RG flows of 
the bi-Yang--Baxter model, Eq.(4.9)  of \cite{Sfetsos:2015nya}.\footnote{Under the redefinitions $(\eta,\tilde\eta,K)\to(2\eta,2\zeta,\frac{1}{8t})$
and for $c^2=-1$ (complex branch).}

\end{enumerate}

\section{More on group spaces}
\label{more.group}

In this section we study the strong integrability of the most general $\l$-deformation 
with $n$ commuting current algebras at different levels \cite{Georgiou:2018gpe}.\footnote{
A perhaps related class of integrable $\sigma$-models was recently constructed in \cite{Delduc:2018hty,Delduc:2019bcl}, based on \cite{Vicedo:2017cge}, following a Hamiltonian approach. The precise relation to the models of \cite{Georgiou:2018gpe} remains to be elucidated.} 
The action of this integrable model is given by Eqs.(2.13), (3.2) in \cite{Georgiou:2018gpe}. 
In particular, for small values of the deformation parameters, the action describes mutual and self-interactions
\be
S_{k_i,\l}=\sum_{i=1}^n S_{\text{WZW},k_i}(\frak{g}_i)+\sum_{i,j=1}^n\sqrt{k_ik_j}\int_{\del B}\text{d}^2\s\,
\text{Tr}\left[J_{i+}\left(\l^{-1}\right)^{-1}_{ij} J_{j-}\right]+\cdots\,,
\ee
where $\left(\l^{-1}\right)^{-1}_{ij}\neq\l_{ij}$, since the inverse in $\l^{-1}$ is taken in the group space, whereas 
$\left(\l^{-1}\right)^{-1}_{ij}$ in the space of couplings. Also, the dots denote subleading terms in the  small $\l$-expansion.
Using the Lax matrices, which were given in Eqs.(3.15)-(3.22) of the aforementioned work, we find that the corresponding spatial parts take the Maillet $r/s$-algebra \eqref{rsMaillet} with
\be
\label{rpmost}
\begin{split}
&r_{+zw}=-\frac{2zw(z-1/\sqrt{k_1})}{(z-w)(d+d_1(z))}\,,\quad  r_{-zw}=-r_{+wz}\,,\\
&\hat  r_{+zw}=\frac{2zw(z-1/\sqrt{k_n})}{(z-w)(\hat d+\hat d_1(z))}\,,\quad \hat r_{-zw}=-\hat r_{+wz}\,,
\end{split}
\ee
where $(d,d_1;\hat d,\hat d_1)$ were given in Eqs.(3.15), (3.19), (3.20), (3.22) of \cite{Georgiou:2018gpe}.\footnote{The two coupling case studied in 
\cite{Georgiou:2018hpd}, can be obtained as a special case of \eqref{rpmost} and the remaining equations of this section for $n=2$.}
As a consi\-stency check we have verified that the mYB equations \eqref{mYB} are satisfied.

To analyze the underlying symmetries we rewrite \eqref{rpmost} in terms of the meromo\-rphic twist functions $(\varphi_\l,\widehat\varphi_\l)$ as in \eqref{skfufsiusk}, where
\be
\varphi^{-1}_\l(z^{-1})=-\frac{z-1/\sqrt{k_1}}{d+d_1(z)}\,,\quad \widehat\varphi^{-1}_\l(z^{-1})=\frac{z-1/\sqrt{k_n}}{\hat d+\hat d_1(z)}\,,
\ee
with poles of order one respectively at 
\be
\begin{split}
&\varphi^{-1}_\l(z^{-1})=0\quad\Longrightarrow\quad z_1=\frac{1}{\sqrt{k_1}\l_{11}}\,,\cdots\,,z_{n-1}=\frac{1}{\sqrt{k_{n-1}}\l_{(n-1)1}}\,,\quad z_n=\frac{1}{\sqrt{k_1}}\,,\\
&\widehat\varphi^{-1}_\l(z^{-1})=0\quad\Longrightarrow\quad \hat z_1=\frac{1}{\sqrt{k_n}}\,,\quad  \hat z_2=\frac{1}{\sqrt{k_2}\l_{n2}}\,,\cdots\,, \hat z_n=\frac{1}{\sqrt{k_n}\l_{nn}}\,.
\end{split}
\ee
Evaluating the spatial Lax matrix and $s_{zw}$ at the poles we obtain
\be
\begin{split}
&L_1(z_1)=-J_{1-}\,,\cdots\,, L_1(z_{n-1})=-J_{(n-1)-}\,,\quad L_1(z_n)=J_{1+}\,,\\
&\widehat L_1(\hat z_1)=-J_{n-}\,,\quad \widehat L_1(\hat z_2)=J_{2+}\,,\cdots\,, \widehat L_1(\hat z_n)=J_{n+}\,,\\
&s_{z_1z_1}=\frac{1}{k_1}\,,\cdots\,, s_{z_{n-1}z_{n-1}}=\frac{1}{k_{n-1}}\,,\quad s_{z_nz_n}=-\frac{1}{k_1}\,, \\
&\hat s_{\hat z_1\hat z_1}=\frac{1}{k_n}\,,\quad \hat s_{\hat z_2\hat z_2}=-\frac{1}{k_2}\,,\cdots\,,  \hat s_{\hat z_n\hat z_n}=-\frac{1}{k_n}\,,\\
&s_{z_iz_j}=0=\hat s_{\hat z_i\hat z_j}\, ,\quad {\rm when}\  i\neq j\, .
\end{split}
\ee
Using the above in the Maillet bracket \eqref{rsMaillet}, we readily obtain $n$ commuting copies of the current algebra \eqref{KM2}.

\section*{Acknowledgements}

We would like to thank Beno\^it Vicedo and Marc Magro for useful communications after a first version of this work appeared. 
George Georgiou and Konstantinos Siampos work on this project has received funding from the Hellenic Foundation
for Research and Innovation (HFRI) and the General Secretariat for Research and Technology (GSRT), 
under grant agreement No 15425.

\begin{appendices}

\end{appendices}


\begin{thebibliography}{1}

\bibitem{Sklyanin:1980ij}
  E.~K.~Sklyanin,
  {\it Quantum version of the method of inverse scattering problem},\\
\href{http://link.springer.com/article/10.1007\%2FBF01091462} {J. Sov. Math. {\bf 19} (1982) 1546},
   Zap.\ Nauchn.\ Semin.\  {\bf 95} (1980) 55.
  
  \bibitem{Maillet:1985ek}
  J.~M.~Maillet,
 {\it New Integrable Canonical Structures in Two-dimensional Models},\\
 \href{http://www.sciencedirect.com/science/article/pii/0550321386903652}{Nucl. Phys. {\bf B269} (1986) 54.}

\bibitem{Maillet:1985ec}
  J.~M.~Maillet,
  {\it Hamiltonian Structures for Integrable Classical Theories From Graded Kac-moody Algebras},
 \href{http://www.sciencedirect.com/science/article/pii/037026938691289X}{Phys. Lett. {\bf B167} (1986) 401.}
 

  \bibitem{Sfetsos:2013wia}
  K.~Sfetsos,
  {\it Integrable interpolations: From exact CFTs to non-Abelian T-duals,}\\
  Nucl. Phys. {\bf B880} (2014) 225,
 \href{http://arxiv.org/abs/1312.4560}{arXiv:1312.4560 [hep-th].}
 
  \bibitem{Balog:1993es}
  J.~Balog, P.~Forgacs, Z.~Horvath and L.~Palla,
 {\it A new family of su(2) symmetric integrable sigma models,}
  Phys. Lett. {\bf B324} (1994) 403, \href{http://arxiv.org/abs/hep-th/9307030}{hep-th/9307030}.
 
 
 \bibitem{Hollowood:2014rla} 
  T.~J.~Hollowood, J.~L.~Miramontes and D.~M.~Schmidtt,
  {\it Integrable Deformations of Strings on Symmetric Spaces},
  JHEP {\bf 1411}, (2014) 009,
  \href{https://arxiv.org/abs/1407.2840}{arXiv:1407.2840 [hep-th].}
  

 \bibitem{Hollowood:2014qma}
  T.~J.~Hollowood, J.~L.~Miramontes and D.~M.~Schmidtt,
  {\it An Integrable Deformation of the $AdS_5 \times S^5$ Superstring},\hfill\break
  J.\ Phys.\ {\bf A47}, no.49 (2014)  495402,
  \href{https://arxiv.org/abs/1409.1538}{arXiv:1409.1538 [hep-th].}
  
 
 
  \bibitem{Itsios:2014vfa} 
  G.~Itsios, K.~Sfetsos, K.~Siampos and A.~Torrielli,
  {\it The classical Yang-Baxter equation and the associated Yangian symmetry of gauged WZW-type theories},\\
  Nucl.\ Phys.\ {\bf B889} (2014) 64,
  \href{https://arxiv.org/abs/1409.0554}{arXiv:1409.0554 [hep-th].}
 

 
 \bibitem{Rajeev:1988hq}
  S.~G.~Rajeev,
  {\it Nonabelian Bosonization Without Wess-zumino Terms. 1. New Current Algebra},
\href{http://www.sciencedirect.com/science/article/pii/0370269389915281}{Phys. Lett. {\bf B217} (1989) 123}.

    \bibitem{Bowcock:1988xr}
  P.~Bowcock,
  {\it Canonical Quantization of the Gauged Wess--Zumino Model},\hfill\break
 \href{http://www.sciencedirect.com/science/article/pii/0550321389903878}{Nucl. Phys. {\bf B316} (1989) 80.}


\bibitem{Sfetsos:2014lla} 
  K.~Sfetsos and K.~Siampos,
  {\it The anisotropic $\lambda$-deformed $SU(2)$ model is integrable},
  Phys.\ Lett.\ {\bf B743} (2015) 160,
  \href{https://arxiv.org/abs/1412.5181}{arXiv:1412.5181 [hep-th].}
  
 
  
  \bibitem{Hollowood:2015dpa} 
  T.~J.~Hollowood, J.~L.~Miramontes and D.~M.~Schmidtt,
  {\it S-Matrices and Quantum Group Symmetry of k-Deformed Sigma Models},\hfill\break
  J.\ Phys.\ {\bf A49}, no. 46 (2016) 465201,
  \href{https://arxiv.org/abs/1506.06601}{arXiv:1506.06601 [hep-th].}
  
  \bibitem{Faddeev:1985qu} 
  L.~D.~Faddeev and N.~Y.~Reshetikhin,
  {\it Integrability of the Principal Chiral Field Model in (1+1)-dimension},
  \href{https://www.sciencedirect.com/science/article/pii/0003491686902010?via\%3Dihub}{Annals Phys.\  {\bf 167} (1986) 227.}

\bibitem{Georgiou:2016iom} 
  G.~Georgiou, K.~Sfetsos and K.~Siampos,
  {\it All-loop correlators of integrable $\l$-deformed $\s$-models},
  Nucl.\ Phys.\ {\bf B909} (2016) 360,
  \href{https://arxiv.org/abs/1604.08212}{arXiv:1604.08212 [hep-th].}
  
  
  
  \bibitem{Curtright:1994be}
  T.~Curtright and C.~K.~Zachos,
  {\it Currents, charges, and canonical structure of pseudodual chiral models},
  Phys. Rev. {\bf D49} (1994) 5408,
  \href{http://arxiv.org/abs/hep-th/9401006}{hep-th/9401006}.

\bibitem{Lozano:1995jx}
  Y.~Lozano,
  {\it Non-Abelian duality and canonical transformations},\hfill\break
  Phys. Lett. {\bf B355} (1995) 165,
 \href{http://arxiv.org/abs/hep-th/9503045}{hep-th/9503045}.

  
  \bibitem{Sfetsos:1996pm}
  K.~Sfetsos,
  {\it Non-Abelian duality, parafermions and supersymmetry},\hfill\break
  Phys. Rev. {\bf D54} (1996) 1682,
  \href{http://arxiv.org/abs/hep-th/9602179}{hep-th/9602179}.


  
  \bibitem{Georgiou:2017jfi} 
  G.~Georgiou and K.~Sfetsos,
  {\it Integrable flows between exact CFTs},\hfill\break
  JHEP {\bf 1711} (2017) 078,
  \href{https://arxiv.org/abs/1707.05149}{arXiv:1707.05149 [hep-th].}
  

  \bibitem{Delduc:2019bcl}
  F.~Delduc, S.~Lacroix, M.~Magro and B.~Vicedo,
  {\it Assembling integrable $\sigma$-models as affine Gaudin models},
  JHEP {\bf 1906} (2019) 017,
  \href{https://arxiv.org/abs/1903.00368}{arXiv:1903.00368 [hep-th].}

  
\bibitem{Georgiou:2017oly} 
  G.~Georgiou, K.~Sfetsos and K.~Siampos,
  {\it Double and cyclic $\lambda$-deformations and their canonical equivalents},
  Phys.\ Lett.\ {\bf B771} (2017) 576,
  \href{http://arxiv.org/abs/arXiv:1704.07834}{arXiv:1704.07834 [hep-th]}.
  
 
 
 
\bibitem{Georgiou:2016urf} 
  G.~Georgiou and K.~Sfetsos,
  {\it A new class of integrable deformations of CFTs},\hfill\break
  JHEP {\bf 1703} (2017) 083, 
  \href{http://arxiv.org/abs/arXiv:1612.05012}{arXiv:1612.05012 [hep-th]}.
  

  \bibitem{Magro:2008dv} 
  M.~Magro,
  {\it The Classical Exchange Algebra of $\text{AdS}_5 \times S^5$}, \hfill\break
  JHEP {\bf 0901} (2009) 021,
  \href{https://arxiv.org/abs/0810.4136}{arXiv:0810.4136 [hep-th].}
  
   \bibitem{Vicedo:2009sn} 
  B.~Vicedo,
  {\it Hamiltonian dynamics and the hidden symmetries of the $\text{AdS}_5 \times S^5$ superstring},
  JHEP {\bf 1001} (2010) 102,
  \href{https://arxiv.org/abs/0910.0221}{arXiv:0910.0221 [hep-th].}
  


  
  \bibitem{Mikhailov:2007eg} 
  A.~Mikhailov and S.~Sch\"afer-Nameki,
  {\it Algebra of transfer-matrices and Yang-Baxter equations on the string worldsheet in  $\text{AdS}_5 \times S^5$},\hfill\break
  Nucl. Phys. {\bf B802} (2008) 1,
  \href{https://arxiv.org/abs/0712.4278}{arXiv:0712.4278 [hep-th].}
  
  
  
    \bibitem{Bardakci:1990ad}
  K.~Bardakci, M.~J.~Crescimanno and S.~Hotes,
  {\it Parafermions from nonabelian coset models},
 \href{http://www.sciencedirect.com/science/article/pii/055032139190332R}{Nucl. Phys. {\bf B349} (1991) 439}.
 
  \bibitem{Bardacki:1990wj}
  K.~Bardakci, M.~J.~Crescimanno and E.~Rabinovici,
  {\it Parafermions From Coset Models},
  \href{http://www.sciencedirect.com/science/article/pii/055032139090365K?via\%3Dihub}{Nucl.\ Phys.\ {\bf B344} (1990) 344.}
  
   \bibitem{Vicedo:2010qd} 
  B.~Vicedo,
  {\it The classical R-matrix of AdS/CFT and its Lie dialgebra structure},\hfill\break
  Lett. Math.\ Phys.  {\bf 95} (2011) 249,
  \href{https://arxiv.org/abs/1003.1192}{arXiv:1003.1192 [hep-th].}
  
  \bibitem{Delduc:2012qb} 
  F.~Delduc, M.~Magro and B.~Vicedo,
  {\it Alleviating the non-ultralocality of coset sigma models through a generalized Faddeev-Reshetikhin procedure},\hfill\break
  JHEP {\bf 1208} (2012) 019,
  \href{https://arxiv.org/abs/1204.0766}{arXiv:1204.0766 [hep-th].}

\bibitem{Sfetsos:2017sep} 
  K.~Sfetsos and K.~Siampos,
  {\it Integrable deformations of the $G_{k_1} \times G_{k_2}/G_{k_1+k_2}$ coset CFTs},
  Nucl.\ Phys.\  {\bf B927} (2018) 124,
  \href{https://arxiv.org/abs/1710.02515}{arXiv:1710.02515 [hep-th].}
  
  \bibitem{Delduc:2015xdm} 
  F.~Delduc, S.~Lacroix, M.~Magro and B.~Vicedo,
  {\it On the Hamiltonian integrability of the bi-Yang-Baxter sigma-model},
  JHEP {\bf 1603} (2016) 104,
  \href{https://arxiv.org/abs/1512.02462}{arXiv:1512.02462 [hep-th].}

\bibitem{Delduc:2013fga} 
  F.~Delduc, M.~Magro and B.~Vicedo,
  {\it On classical $q$-deformations of integrable sigma-models},
  JHEP {\bf 1311}, (2013) 192,
  \href{https://arxiv.org/abs/1308.3581}{arXiv:1308.3581 [hep-th].}
  
  
  
  
  \bibitem{Vicedo:2015pna}
  B.~Vicedo,
  {\it Deformed integrable $\sigma$-models, classical $R$-matrices and classical exchange algebra on Drinfel'd doubles},\hfill\break
  J. Phys. A: Math. Theor. {\bf 48} (2015) 355203,
 \href{http://arxiv.org/abs/1504.06303}{arXiv:1504.06303 [hep-th]}.
  
  \bibitem{Hoare:2015gda}
  B.~Hoare and A.~A.~Tseytlin,
  {\it On integrable deformations of superstring sigma models related to $AdS_n \times S^n$ supercosets},\hfill\break
  {Nucl.\ Phys.\ {\bf B897} (2015) 448,}
  \href{http://arxiv.org/abs/1504.07213}{arXiv:1504.07213 [hep-th].}


\bibitem{Sfetsos:2015nya}
  K.~Sfetsos, K.~Siampos and D.C.~Thompson,
 {\it Generalised integrable $\l$- and $\eta$-deformations and their relation},\hfill\break
  Nucl. Phys. {\bf B899} (2015) 489,
  \href{http://arxiv.org/abs/1506.05784}{arXiv:1506.05784 [hep-th].}


\bibitem{Klimcik:2015gba}
C. Klim\v c\'\i k,
  {\it $\eta$ and $\lambda$ deformations as ${\cal E}$-models},\hfill\break
   Nucl. Phys. {\bf B900} (2015) 259,
  \href{http://arxiv.org/abs/1508.05832}{arXiv:1508.05832 [hep-th].}
  
  \bibitem{Klimcik:2016rov}
C. Klim\v c\'\i k,
  {\it Poisson--Lie T-duals of the bi-Yang--Baxter models},\\
  Phys.\ Lett.\ {\bf B760} (2016) 345,
  \href{https://arxiv.org/abs/1606.03016}{arXiv:1606.03016 [hep-th]}.
  
    \bibitem{Georgiou:2018gpe}
  G.~Georgiou and K.~Sfetsos,
  {\it The most general $\lambda$-deformation of CFTs and integrability},
  JHEP {\bf 1903} (2019) 094,
  \href{https://arxiv.org/abs/1812.04033}{arXiv:1812.04033 [hep-th].}
  

  \bibitem{Delduc:2018hty} 
  F.~Delduc, S.~Lacroix, M.~Magro and B.~Vicedo,
  {\it Integrable Coupled $\sigma$ Models},\\
  Phys.\ Rev.\ Lett.\  {\bf 122}, no. 4, (2019) 041601,
  \href{https://arxiv.org/abs/1811.12316}{arXiv:1811.12316 [hep-th].}
  
  
  \bibitem{Vicedo:2017cge} 
  B.~Vicedo,
  {\it On integrable field theories as dihedral affine Gaudin models},\\
  International Mathematics Research Notices {\bf rny128} (2017),\\
  \href{https://arxiv.org/abs/1701.04856}{arXiv:1701.04856 [hep-th].}
  
  
  
  
  \bibitem{Georgiou:2018hpd} 
  G.~Georgiou and K.~Sfetsos,
  {\it Novel all loop actions of interacting CFTs: Construction, integrability and RG flows},
  Nucl. Phys. {\bf B937} (2018) 371,
  \href{https://arxiv.org/abs/1809.03522}{arXiv:1809.03522 [hep-th].}



\end{thebibliography}
\end{document}